\documentclass[12pt]{article}

\usepackage{graphicx}

\usepackage{amssymb}
\usepackage{amsmath}

\usepackage{amsbsy}

\usepackage{caption}

\begin{document}%



\title{\bf Approximate formula for total cross section for moderately small eikonal function}

\author{
A.V. Kisselev\thanks{Electronic address:
alexandre.kisselev@ihep.ru} \\
{\small A.A. Logunov Institute for High Energy Physics, NRC
``Kurchatov Institute'',} \\
{\small 142281, Protvino, Russian Federation} }

\date{}

\maketitle

\begin{abstract}
The eikonal approximation for the total cross section for the
scattering of two unpolarized particles is studied. The approximate
formula in the case when the eikonal function $\chi(b)$ is
moderately small, $|\chi(b)| \lesssim 0.1$, is derived. It is shown
that the total cross section is given by the series of multiple
improper integrals of the Born amplitude $A_B$. Its advantage
compared to standard eikonal formulas is that the integrals contain
no rapidly oscillating Bessel functions. Two theorems which allow
one to relate large--$b$ behavior of $\chi(b)$ with analytical
properties of the Born amplitude are proved. Several examples of
these theorems are given. To check the efficiency of the main
formula, it is applied for numerical calculations of the total cross
section for a number of particular expressions of $A_B$. Only those
Born amplitudes are chosen which result in moderately small eikonal
functions and lead to the correct asymptotics of  $\chi(b)$. The
numerical calculations show that our formula approximates the total
cross section with the relative error of $\mathrm{O} (10^{-5})$,
provided that the first three non-zero terms in it are taken into
account.
\end{abstract}




\section{Introduction} %
\label{sec:int}

In potential scattering the eikonal approximation is applied when a
scattering angle is small, and energy of incoming particle is much
larger than a ``strength'' of an interaction
potential~\cite{Moliere}, \cite{Glauber}. In quantum field theory
the use of the eikonal approximation is justified if
\begin{equation}\label{eik_kinematic}
-t/s \ll 1 \;,
\end{equation}
where $s$, $t$ are the Mandelstam variables. In perturbation theory
the eikonal representation was studied in \cite{Cheng}. It can be
derived in the framework of quasipotential approach for small
scattering angles and smooth quasipotentials~\cite{Logunov}. In the
Regge approach \cite{Collins}, the eikonalization corresponds to the
summation of contributions of multi-reggeon exchanges.

In the eikonal approximation, the total cross section for a
scattering of unpolarized particles is given by the formula
\begin{equation}\label{eik_cs}
\sigma_{\mathrm{tot}}(s) = 4\pi \!\! \int\limits_0^{\infty} \!db b
\left\{ 1 - \exp[-\mathrm{Im}\chi (s,b)] \cos[\mathrm{Re}\chi (s,b)]
\right\} ,
\end{equation}
where $b$ is the impact parameter. The eikonal function $\chi(s,b)$
is related to the Born amplitude $A_B(s,t)$ by the Fourier-Bessel
transformation
\begin{equation}\label{eikonal}
\chi(s, b) = \frac{1}{4\pi s} \int\limits_0^{\infty} \! dq q J_0(q
b) \, A_{\mathrm{B}} (s, q) \;.
\end{equation}
Here and in what follows, $q = \sqrt{-t}$. In its turn, the Born
amplitude is defined via eikonal function as
\begin{equation}\label{Born_amplitude}
A_{\mathrm{B}}(s, q)  = 4\pi s \!\int\limits_0^{\infty} \! db b \,
J_0(q b) \, \chi(s, b)  \;.
\end{equation}
Both the eikonal function and Born amplitude are dimensionless
quantities.

As we see from Eqs.~\eqref{eik_cs}-\eqref{Born_amplitude}, in order
to calculate the total cross section, one has to deal with integrals
containing \emph{rapidly oscillating} Bessel functions. The
difficulties in evaluating integrals over half-line involving
products of Bessel functions were considered in Refs.~\cite{Lucas},
\cite{Deun}. Note that integrals with the products of the Bessel
functions are very difficult to evaluate numerically because of
their poor convergence and oscillatory nature.

As it was recently shown in \cite{Kisselev:16}, this problem can be
solved for moderately small eikonal functions. We call the eikonal
function $\chi(s, b)$ \emph{moderately small} if $|\chi(s, b)|
\lesssim 0.1$, but the inequality $|\chi(s, b)| \ll 1$
is not implied.%
\footnote{The \emph{small} eikonal function, $|\chi(s, b)| \ll 1$,
is not so interesting, since corresponding formulas are considerably
simplified in this case.}
In Ref.~\cite{Kisselev:16} approximate formula for a
\emph{scattering amplitude} was derived which contains no Bessel
functions, and, hence, no rapidly oscillating integrands. In the
present paper we derive analogous approximate formula for the
\emph{total cross section} \eqref{eik_cs} which can be used for
numerical calculations of $\sigma_{\mathrm{tot}}$.

The paper is organized as follows. In the next section a series
expansion of the total cross section is derived. In
Section~\ref{sec:exp} the eikonal function at large values of the
impact parameter is analyzed. Two relevant theorems for the Hankel
transform of order zero are proved in Section \ref{sec:Hankel}. In
the last section our theoretical formula is applied for numerical
calculations of $\sigma_{\mathrm{tot}}$ for a number of particular
expressions for Born amplitude in order to study its efficiency.
Some relevant formulas are collected in Appendices~A, B, C and D.



\section{Series expansion of total cross section} %
\label{sec:eqs}

We assume that $|\chi(s, b)|$ is moderately small. Let us start from
the expansion of the integrand in the r.h.s. of Eq.~\eqref{eik_cs}
\begin{align}\label{integrand_expansion}
&1 - e^{-\mathrm{Im}\chi} \cos(\mathrm{Re}\chi) = \mathrm{Im}\chi +
\frac{1}{2} \, [\mathrm{Re}\chi^2 - \mathrm{Im}\chi^2] - \frac{1}{6}
\, \mathrm{Im}\chi [3 \mathrm{Re}\chi^2 - \mathrm{Im}\chi^2 ]
\nonumber \\
&+ \frac{1}{24} \, [6\,\mathrm{Re}\chi^2 \mathrm{Im}\chi^2 -
\mathrm{Re}\chi^4 - \mathrm{Im}\chi^4]
\nonumber \\
&- \frac{1}{120} \, \mathrm{Im}\chi [10\,\mathrm{Re}\chi^2
\mathrm{Im}\chi^2 - 5\,\mathrm{Re}\chi^4 - \mathrm{Im}\chi^4]
\nonumber \\
&- \frac{1}{720} [15\,\mathrm{Re}\chi^4 \mathrm{Im}\chi^2 -
15\,\mathrm{Re}\chi^2 \mathrm{Im}\chi^4 - \mathrm{Re}\chi^6 +
\mathrm{Im}\chi^6] + \ldots \;.
\end{align}
We omitted higher terms in \eqref{integrand_expansion}. As we will
see in Section~\ref{sec:num}, an account of only four terms in this
expansion is enough to approximate $\sigma_{\mathrm{tot}}$ with a
very small relative error.

Correspondingly, we obtain the following expansion of the total
cross section
\begin{equation}\label{cs_expansion}
\sigma_{\mathrm{tot}}(s) = \sigma_1(s) + \ldots +  \sigma_6(s) +
\ldots  \;.
\end{equation}
We find from Eqs.~\eqref{eik_cs}--\eqref{cs_expansion}
\begin{equation}\label{sigma_1}
\sigma_1(s) = \frac{\mathrm{Im}A_B(s,0)}{s} \;,
\end{equation}
\begin{equation}\label{sigma_2}
\sigma_2(s) = \frac{1}{2^3\pi s^2} \!\int\limits_0^{\infty} \!dq q
\{ [\mathrm{Re}A_B(s,q)]^2 - [\mathrm{Im}A_B(s,q)]^2 \} \;.
\end{equation}
In deriving \eqref{sigma_2}, we used Eq.~\eqref{two_Bessels} from
Appendix~B. We assume that $\mathrm{Im}A_B(s,0)$ exists.%
\footnote{This is not a case in theories with massless particles.}
Note, the optical theorem says that
\begin{equation}\label{sigma_1_full}
\sigma_{\mathrm{tot}}(s) = \frac{\mathrm{Im}A(s,0)}{s} \;,
\end{equation}
where $A(s,q)$ is the \emph{full} scattering amplitude. The
expressions for other four terms in \eqref{cs_expansion} are given
in Appendix~A.

Provided the Born amplitude $A_B(s,q)$ is known at fixed $s$ as a
function of $q$, the formulas presented in this section and
Appendix~A enable us to numerically calculate the total cross
section for the same energy (see numerical examples in
Section~\ref{sec:num}).

\section{Eikonal function at large values of impact parameter} %
\label{sec:exp}

The main goal of this section is to study an asymptotics of the
eikonal function as $b \rightarrow \infty$. Let us start from the
partial wave expansion of the scattering amplitude
\begin{equation}\label{par_wave_exp}
A(s,t) = 16\pi \sum_{l=0}^\infty (2l+1) A_l(s) P_l(\cos\theta) \;.
\end{equation}
We assume that $s \gg m^2$, where $t_0 = m^2$ is a nearby
singularity in variable $t$. The $s$-channel partial waves
exponentially decrease as $l$ grows \cite{Collins}
\begin{equation}\label{part_wave_ampl_bound}
A_l(s)\big|_{l,s \rightarrow \infty} = f(s) \exp(- 2 l m/\sqrt{s})
\;.
\end{equation}
We have at small $\theta$
\begin{equation}\label{t_s_theta}
-t = \frac{s}{2} (1 - \cos\theta)\big|_{\theta \ll 1} = \frac{s
\,\theta^2}{4} \;,
\end{equation}
\begin{equation}\label{theta_vs_l}
\theta = \frac{2q}{\sqrt{s}} = \frac{bq}{l} \;.
\end{equation}
Correspondingly,
\begin{equation}\label{Pl}
P_l(\cos\theta)\big|_{\theta \ll 1} = P_l \!\left (\cos \frac{bq}{l}
\right)\!\Big|_{l \gg 1} .
\end{equation}

Let $P_\nu^\mu(z)$ be the Legendre function of the first kind
\cite{Bateman_vol_1}. If $\mu, x$ are fixed and $\nu \rightarrow
\infty$ through real positive values, then (see Eq.~9.1.71 in
\cite{Abramowitz})
\begin{equation}\label{Legendre_limit}
\lim_{\nu \rightarrow \infty} \left[ \nu^\mu P_\nu^{-\mu} \!\left
(\cos \frac{x}{\nu} \right) \right] = J_\mu (x) \quad (x>0) \;,
\end{equation}
The case $\nu = n$, $\mu = -m$ was considered in \cite{Whittaker}
(see Section~17.4, Example~1). The special case $\nu = n$, $\mu =
0$, was studied in \cite{Mehler} (Mehler--Heine formula).%
\footnote{A generalization of the Mehler--Heine formula to Jacobi
polynomials $P^{\alpha,\beta}(x)$ is given by G.~Szeg\H{o}
\cite{Szego}.}
The less complicated derivation of formula \eqref{Legendre_limit} is
presented in Appendix~C. Thus, the summation over large $l$ in
\eqref{par_wave_exp} can be replaced by integration over large $b$
\begin{equation}\label{large_l}
16\pi \!\sum_{l=l_0 \gg 1}^\infty (2l+1) A_l(s) P_l(\cos\theta) =
8\pi s \!\!\int\limits_{b_0(s)}^\infty \!\!db b \,A(b,s)J_0(q b) \;,
\end{equation}
where $b_0(s) = l_0 \sqrt{s}/2$. On the other hand, we have
\begin{equation}\label{eik_ampl}
A(s,t) = 4\pi i s \!\!\int\limits_0^\infty \!db b \left[ 1 - e^{i
\chi (b,s)} \right] \!J_0(q b) \;.
\end{equation}
By comparing Eqs.~\eqref{large_l} and \eqref{eik_ampl}, using
Eq.~\eqref{part_wave_ampl_bound}, we get
\begin{equation}\label{eik_bound}
|\chi(b, s)|\big|_{\genfrac{}{}{0pt}{}{s \gg m^2}{\!\!b m \gg 1}} =
2f(s)\exp(-b m) \;.
\end{equation}
Note that the asymptotics of the eikonal function is factorized.

Let us consider one example. Suppose that at small $t$ the Born
amplitude is given by $t$-channel exchange
\begin{equation}\label{small_t}
A_{\mathrm{Born}}(s,t)\Big|_{-t \lesssim m^2} \sim \frac{1}{m^2 - t}
\;,
\end{equation}
while it exponentially decreases as $|t|$ grows
\begin{equation}\label{large_t}
A_{\mathrm{Born}}(s,t)\Big|_{-t \gg m^2} \sim e^{ct} \;,
\end{equation}
where $c>0$. If we put
\begin{equation}\label{Born_ampl}
A_{\mathrm{Born}}(s,t) = g(s) \frac{e^{ct}}{m^2 - t} \;,
\end{equation}
we come to the following expression for the eikonal function
\cite{Kisselev:18}
\begin{align}\label{int_2}
\chi(b,s) &= \frac{g(s)}{4\pi s} \int\limits_0^\infty \!\frac{q}{m^2
+ q^2} \,e^{-q^2 c} \,J_0(q b) \,dq = G(s) \Bigg[ e^{m^2 c} K_0(b m)
\nonumber \\
&- \frac{2c}{b^2} \,e^{-b^2/4c} \sum_{p=0}^\infty \left( \frac{2m
c}{b} \right)^{\!\!2p} \!{}_2F_0 \!\left(p+1, p+1; -\frac{4c}{b^2}
\right) \Bigg] ,
\end{align}
where $G(s) = g(s)/(4\pi s)$, $K_0(z)$ is the modified Bessel
function of the second kind (Macdonald function)
\cite{Bateman_vol_2}, and $_2F_0 (a,b;z)$ is the generalized
hypergeometric function \cite{Bateman_vol_1}. As a result, we find the asymptotics%
\footnote{This asymptotics was also derived in Ref.~\cite{Frenzen},
see formula (4.20) there.}
\begin{equation}\label{chi_asym}
\chi(b,s) \Big|_{b m \gg 1} = G(s) \sqrt{\frac{\pi}{2b m}} \,e^{-b m
+ m^2 c} \left[1 + \mathrm{O}((bm)^{-1}) \right] ,
\end{equation}
in full agreement with Eq.~\eqref{eik_bound}.

This asymptotics is also in agreement with general results of
Ref.~\cite{Frenzen}. To recall them, consider the following zero
order Hankel transform
\begin{equation}\label{Frenzen_int}
I_f(w) = \int\limits_0^\infty \!\!x e^{-x^2} \!f(x^2) J_0(w x) \,dx
\;,
\end{equation}
where $f(z)$ is a \emph{meromorphic} function such that $f(z^2)$ has
a finite number of poles at $a_1, \ldots , a_k$ in the upper
half-plane and satisfies
\begin{equation}\label{f_condition}
|f(z)| \leqslant M e^{\xi \mathrm{Re z + \eta |\mathrm{Im z}|}} \;,
\end{equation}
with $M > 0$, $\eta \geqslant 0$, $\xi < 1$. Then according to
Theorem~2 in \cite{Frenzen}
\begin{equation}\label{f_asymp_2}
I_f(w) = \mathrm{O} \left( e^{- \delta w} \right) ,
\end{equation}
as $w \rightarrow \infty$, where $\delta = \min(\mathrm{Im}\,a_1,
\ldots, \mathrm{Im}\,a_k)$. In our case \eqref{Born_ampl}, $f(z^2) =
1/[(m c)^2 + z^2]$, $w= b/c$, $\delta = m c$.

Note that the $t$-channel exchange \eqref{small_t} itself leads to
the same exponentially decreasing asymptotics
\begin{equation}\label{ampl_one_part_exchage}
\int\limits_0^\infty \frac{q}{m^2 + q^2} \,J_0(b q) \,dq =  K_0(b m)
\Big|_{b m \gg 1} = \sqrt{\frac{\pi}{2b m}} \,e^{-b m} \left[1 +
\mathrm{O}((b m)^{-1}) \right] .
\end{equation}

It is also necessary to mention another important theorem in
\cite{Frenzen}. It is formulated as follows. Let $f(z)$ in
\eqref{Frenzen_int} be an \emph{entire} function satisfying
condition \eqref{f_condition} for all sufficiently large $|z|$.
Define
\begin{equation}\label{alpha}
\alpha = 1 - \xi + \eta^2/(1 - \xi) \;.
\end{equation}
Then Theorem~1 in \cite{Frenzen} says that
\begin{equation}\label{f_asymp_1}
I_f(w) = \mathrm{O} \!\left( e^{-w^2/\alpha}\right) ,
\end{equation}
as $w \rightarrow \infty$.

\section{Two theorems for zero order Hankel transform} %
\label{sec:Hankel}

In the light of the above, we formulate and prove two new theorems
for the Hankel transform of order zero of an \emph{even} function
\begin{equation}\label{Frenzen_int_mod}
J_g(w) = \int\limits_0^\infty \!\!x \,\!g(x^2) J_0(w x) \,dx \;.
\end{equation}
Recall that the eikonal function and Born amplitude are related by
this transform \eqref{eikonal}.

\textsc{Theorem~1}. Let $g(z)$ be a \emph{meromorphic} function such
that $g(z^2)$ has a finite number of poles in the upper half-plane
at $c_1, \ldots , c_k$, and let $g(z)$ satisfies the grows condition
of the form
\begin{equation}\label{f_condition_mod}
|g(z)| = \mathrm{O}(|z|^{-c}) \;, \quad c > 1/4 \;,
\end{equation}
for all sufficiently large $|z|$. Then
\begin{equation}\label{Jg_I}
J_g(w) = \pi i \sum_{i=1}^k \mathrm{Res} \!\left[z g(z^2)
H_0^{(1)}(wz); c_i \right] ,
\end{equation}
where $H_0^{(1)}(z)$ is the Hankel function of the first kind
\cite{Bateman_vol_2}.

\begin{figure}[htb]
\begin{center}
\includegraphics[width=8cm]{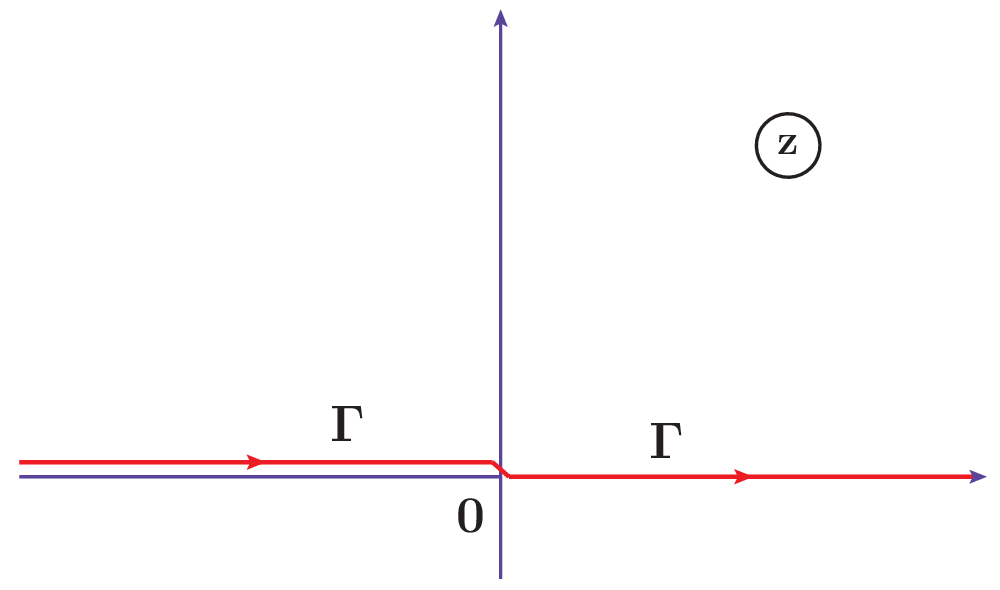}
\caption{The contour $\Gamma$ in integral \eqref{Jg_contour_I}.}
\label{fig:contour_Gamma}
\end{center}
\end{figure}

\textsc{Proof}. Let us use the relation (see Eq.~3.62(5) in
\cite{Watson})
\begin{equation}\label{Bessel_to_Hankel}
J_0(z) = \frac{1}{2} \left[ H_0^{(1)}(z) - H_0^{(1)}(z e^{i\pi})
\right] .
\end{equation}
The Hankel function $H_0^{(1)}(z)$ has a logarithmic branch point at
$z=0$ \cite{Bateman_vol_2}. The branch cut of $H_0^{(1)}(z)$  is
determined by taking the branch of $\ln z$ which is real for
positive $z$. As a result, \eqref{Frenzen_int_mod} takes the form
\begin{equation}\label{Jg_contour_I}
J_g(w) = \frac{1}{2} \int\limits_\Gamma \!\!z \,\!g(z^2) H_0^{(1)}(w
z) dz \;,
\end{equation}
where the contour $\Gamma$ consists of the line along the upper side
of branch cut along the negative real semi-axis and positive real
semi-axis (see Fig.~\ref{fig:contour_Gamma}).
%
Let us consider the integral
\begin{equation}\label{contour_I}
\frac{1}{2} \int\limits_{C} \!\!z \,\!g(z^2) H_0^{(1)}(w z) dz \;,
\end{equation}
where the closed contour $C$ is a sum of the contour $\Gamma$ and
infinite-radius anti-clockwise semi-circle in the upper plane. Since
the Hankel function has the asymptotics \cite{Bateman_vol_2}
\begin{equation}\label{Hankel_asym}
H_\nu^{(1)}(z) = \sqrt{\frac{2}{\pi z}} \,e^{i(4z - 2\nu\pi -
\pi)/4} \left[ 1 + \mathrm{O}(z^{-1}) \right] , \quad -\pi < \arg z
< 2\pi\;,
\end{equation}
the integral along the semicircle is zero. By Cauchy's theorem we
pick up the poles of $g(z^2)$. \textsc{Q.E.D.}

\textsc{Corollary~1}. Under conditions of Theorem~1, the Hankel
transform \eqref{Frenzen_int_mod} decreases exponentially at large
$w$
\begin{equation}\label{exponential_asymp_I}
J_g(w) = \mathrm{O}(e^{-\gamma w}) \;, \quad w \rightarrow \infty
\;,
\end{equation}
where $\gamma = \min (\mathrm{Im} c_1, \ldots, \mathrm{Im} c_k)$. In
order to prove this statement, it is enough to use the relation
\cite{Bateman_vol_2}
\begin{equation}\label{Hnu_to_Knu}
 H_\nu^{(1)} \left( z e^{i\pi/2} \right) = \frac{2}{\pi i} \,e^{i\nu\pi/2}K_\nu(z) \;,
\end{equation}
as well as asymptotics of the Macdonald function
\cite{Bateman_vol_2}
\begin{equation}\label{asym_Knu}
K_\nu(z) \big|_{z \gg 1} = \sqrt{\frac{\pi}{2z}} \,e^{-z}  [1 +
\mathrm{O}(z^{-1})] \;, \quad -\frac{3\pi}{2} < \arg z <
\frac{3\pi}{2} \;.
\end{equation}
We consider two examples of Theorem~1.

\textsc{Example~1}. Let $g(z)$ be a meromorphic function and let
$g(x^2)$ has a simple pole in the upper plane%
\footnote{Here and below we are not interested in poles in the lower
half-plane.}
\begin{equation}\label{simple_pole}
g(x^2) = \frac{1}{x^2 + a^2} \;.
\end{equation}
Then we immediately find from \eqref{Jg_I} that
\begin{equation}\label{example_1}
J_g(w) = \pi i \,\frac{x H_0^{(1)}(w x)}{x+ia} \bigg|_{x=ia} =
K_0(w a) \;,
\end{equation}
where we used Eq.~\eqref{Hnu_to_Knu} with $\nu=0$.

\textsc{Example~2}. Let $g(z)$ be a meromorphic function and let
$g(x^2)$ has a pole of order $n+1$ ($n \geqslant 1$) in the upper
plane
\begin{equation}\label{double_pole}
g(x^2) = \frac{1}{(x^2 + a^2)^{n+1}} \;.
\end{equation}
After changing variable $x = e^{i\pi/2} z$ and taking into account
Eq.~\eqref{Hnu_to_Knu}, we obtain
\begin{align}\label{example_2_1}
&J_g(w) = \frac{\pi i}{n!} \frac{d^n}{dx^n} \frac{x H_0^{(1)}(w
x)}{(x+ia)^{n+1}} \bigg|_{x=ia} = (-1)^n \frac{2}{n!}
\frac{d^n}{dz^n} \frac{z
K_0(w z)}{(z+a)^{n+1}} \bigg|_{z=a} \nonumber \\
&= \frac{(-1)^n}{(2a)^n n!} \sum_{k=0}^{n-1} \binom{n}{k}
\frac{(-1)^k}{2^k} \,\frac{\Gamma(n+k)(n-k)}{\Gamma(n+1)}
\,\frac{1}{z^{n+k}}\frac{d^{n-k}}{dz^{n-k}} \,K_0(w z) \big|_{z=a} ,
\end{align}
where $\binom{n}{k}$ is the binomial coefficient. With the use of
formula \eqref{diff_n} derived in Appendix~D and relation
\cite{Bateman_vol_2}
\begin{equation}\label{dif_Knu}
\left( \frac{1}{z} \frac{d}{dz}\right)^{\!\!m} \!K_0(z) = (-1)^m
z^{-m} K_m(z) \;, \quad m = 1,2,3, \ldots \;,
\end{equation}
we find from \eqref{example_2_1}
\begin{equation}\label{example_2_2}
J_g(w) = \frac{(-1)^n}{(2a)^n n!} \left( \frac{1}{z}
\frac{d}{dz}\right)^{\!\!n} \!\!K_0(w z) \big|_{z=a} = \frac{1}{n!}
\left( \frac{w}{2a}\right)^{\!\!n} K_n(w a) \;.
\end{equation}
It is a correct expressions for the zero order Hankel transform of
function \eqref{double_pole} (see Eq.~2.12.4.28 in
\cite{Prudnikov_vol_2}).

\textsc{Theorem~2}. Let $g(z)$ be an \emph{analytic} function such
that $g(z^2)$ has a finite number of branch points in the upper
half-plane at $a_1+ib_1, \ldots, a_k+ib_k$, where $a_i\neq a_j$ for
$i\neq j$, and let $g(z)$ satisfies the grows condition of the form
\begin{equation}\label{g_condition_mod}
|g(z)| = \mathrm{O}(|z|^{-c}) \;, \quad c>1/4 \;,
\end{equation}
for all sufficiently large $|z|$. Then
\begin{equation}\label{Jg_contour_II}
J_g(w) = -\frac{1}{2} \sum_{i=1}^k \int\limits_{C_i} \!\!z  g(z^2)
H_0^{(1)}(wz) \,dz \;,
\end{equation}
where the clockwise contour $C_i$ ($i=1, \ldots, k$) comes from
infinity along the right side of the branch cut $(a_i+ib_i,
a_i+i\infty)$, goes around the branch point $a_i+ib_i$, and then
tends to infinity along the left side of this branch cut. For given
$C_i$, the branch cut of $zg(z^2)$ is determined by taking the
branch of $g(z)$ which is real along the finite section $(a_i, a_i +
i b_i)$ of the imaginary axis.

\textsc{Proof}. Let us complement the contour $\Gamma$ in
\eqref{Jg_contour_I} to a closed contour $C$ in the upper half-plane
\begin{equation}\label{close_contour}
C = \Gamma + \sum_{i=1}^{k} C_i + \sum_{i=1}^{k+1} C_{R_i} \;,
\end{equation}
where $C_{R_i}$ are infinite-radius anti-clockwise contours, while
$C_i$ are clockwise contours along the sides of the branch cuts
$(a_i+ib_i, a_i+i\infty)$. By Cauchy's integral theorem, the
integral of the analytic function $z \,\!g(z^2) H_0^{(1)}(w z)$ in
the upper half-plane along the closed contour $C$ is equal to zero.

Let $I_{C_i}$ ($I_{R_i}$) be an integral of this function along the
contour $C_i$ ($C_{R_i}$). All integrals $I_{R_i}$ are zero due to
condition \eqref{g_condition_mod} and asymptotic properties of the
Hankel function $H_0^{(1)}(z)$ at large $z$ \eqref{Hankel_asym}.
Then formula \eqref{Jg_contour_II} immediately follows from
Eq.~\eqref{Jg_contour_I}, \textsc{Q.E.D.}

\textsc{Corollary~2}. Let the function $g(z^2)$ has a special form
\begin{equation}\label{g_corollary_II}
g(z^2) = (z^2 - z_1^2)^{\alpha_1} \ldots (z^2 - z_k^2)^{\alpha_k}
f(z^2) \;,
\end{equation}
where $z_i = a_i + ib_i$ with $a_i\neq a_j$, $b_i\neq b_j$ for all
$i,j = 1, 2, \ldots, k$, non integer $\alpha_i > -1$, and $f(z^2)$
is a holomorphic function in the upper half-plane. If
\eqref{g_corollary_II} satisfies asymptotic condition
\eqref{g_condition_mod} of Theorem~2, then the Hankel transform
\eqref{Frenzen_int_mod} decreases exponentially at large $w$
\begin{equation}\label{exponential_asymp_II}
J_g(w) = \mathrm{O}(e^{-\lambda w}) \;, \quad w \rightarrow \infty
\;,
\end{equation}
where $\lambda = \min (b_1, \ldots, b_k)$.

\textsc{Proof}. Suppose that $\lambda = b_p$, $1 \leqslant p
\leqslant k$. After changing variables $z = z_p + iv$, we obtain
from \eqref{Jg_contour_II}, \eqref{g_corollary_II} the contribution
of the branch cut $(a_p + ib_p, a_p + i\infty)$
\begin{equation}\label{Jg_asym_corol_II_1}
J_g^{(p)}(w) = \frac{2}{\pi} \int\limits_0^\infty \!\! v^{\alpha_p}
F(v) K_0[w (b_p + v - ia_p)] \,dv \;,
\end{equation}
where the notation
\begin{align}\label{F}
F(v) &= iz (z^2 - z_1^2)^{\alpha_1} \ldots (z^2 -
z_{p-1}^2)^{\alpha_{p-1}} (z^2 - z_{p+1}^2)^{\alpha_{p+1}} \ldots
(z^2 - z_k^2)^{\alpha_k}  \nonumber \\
&\times [-i(z + z_p)]^{\alpha_p} f(z^2) \big|_{z = z_p + iv}
\sin(\pi \alpha_p)
\end{align}
is introduced. The leading part of the asymptotics of $J_g^{(p)}(w)$
is given by
\begin{align}\label{Jg_asym_corol_II_2}
J_g^{(p)}(w)\big|_{w \rightarrow \infty} &= \sqrt{\frac{2}{\pi w(b_p
- ia_p)}} \,e^{-w (b_p - i a_p)} F(0) \!\int\limits_0^\infty
\!\!v^{\alpha_p} e^{-w v} \,dv \nonumber
\\
&= \sqrt{\frac{2}{b_p - ia_p}} \, w^{-(\alpha_p + 3/2)} \,e^{-w (b_p
- i a_p)} \,F(0)\;.
\end{align}
As $w$ goes to infinity, contributions from other cuts are
non-leading with respect to \eqref{Jg_asym_corol_II_2}, because $b_p
< \min (b_1, \ldots b_{p-1},b_{p+1} \ldots b_k)$. \textsc{Q.E.D.}

Now we consider four examples of Theorem~2.

\textsc{Example~3}. Let $g(z)$ be an analytic function in the upper
half-plane and let $g(x^2)$ has an integrable branch point in the
upper half-plane
\begin{equation}\label{g_example_3}
g(x^2) = \frac{1}{\sqrt{1+x^2}} \;.
\end{equation}
It is a particular case of Theorem~2 with $k=1$ $a_1 = 0, b_1=1$,
and there is the branch cut $(i, i\infty)$ in the upper half-plane.%
\footnote{Here and in what follows we are not interested in branch
cuts in the lower half-plane.}
We have (see Eq.~2.12.4.28 in \cite{Prudnikov_vol_2})
\begin{equation}\label{Jg_example_3}
\int\limits_0^\infty \!\frac{x}{\sqrt{1+x^2}} \,J_0(w x) \,dx =
\frac{1}{w} \,e^{-w} \;.
\end{equation}

\begin{figure}[htb]
\begin{center}
\includegraphics[width=10cm]{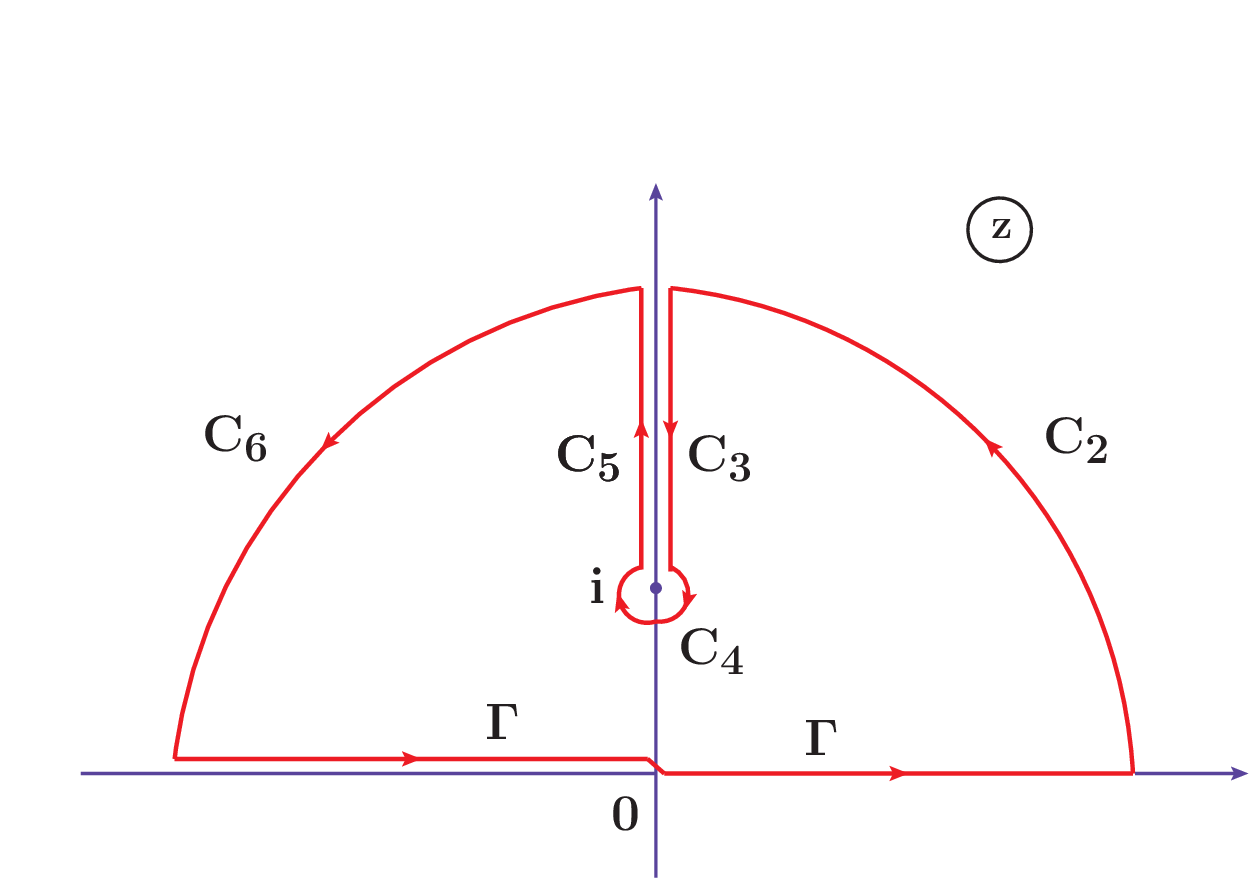}
\caption{The contour of integration for functions
\eqref{g_example_3}, \eqref{g_example_4}.}
\label{fig:contour}
\end{center}
\end{figure}

In Fig.~\ref{fig:contour} a closed contour is shown which is a sum
of the contour $\Gamma$ and contours $C_i$, ($i=2, \ldots 6$). In
particular, the contour $C_4$ is a circle of small radius $r$
centered at $z = i$.
%
According to Theorem~2, $J_g = -(1/2)(I_{C_2} + I_{C_3} + I_{C_4} +
I_{C_5} + I_{C_6})$. Since $I_{C_2} = I_{C_6} = 0$, and $\lim_{r
\rightarrow 0} I_{C_4} = 0$, we get
\begin{equation}\label{example_3}
J_g(w) = -\frac{1}{2}\lim_{r \rightarrow 0}(I_{C_3} + I_{C_5}) \;.
\end{equation}
Let us change variables $z = u e^{i\pi/2}$. The branch cuts $(\pm i,
\pm i\infty)$ in the $z$-plane (see Fig.~\ref{fig:contour}) turn
into the branch cuts $(\pm 1, \pm \infty)$ in the $u$-plane. The
contours $C_3$ and $C_5$ have opposite orientations. On the other
hand, $\sqrt{1-u^2} = -i\sqrt{u^2-1}$ at the upper side of the right
branch cut $(1,\infty)$, while $\sqrt{1-u^2} = i\sqrt{u^2-1}$ at its
lower side. Thus, $I_{C_3} = I_{C_5}$. With the use of relation
\eqref{Hnu_to_Knu} and Eq.~2.16.3.8 in \cite{Prudnikov_vol_2}, we
find that
\begin{equation}\label{example_4_final}
J_g(w) = \frac{2}{\pi} \int\limits_1^\infty \!\frac{u}{\sqrt{u^2-1}}
\, K_0(w u) \,du = \frac{1}{w} \,e^{-w} ,
\end{equation}
in full agreement with Eq.~\eqref{Jg_example_3} and Corollary~2.

\textsc{Example~4}. Let $g(z)$ be an analytic function in the upper
half-plane and let $g(x^2)$ has a non-integrable branch point there
\begin{equation}\label{g_example_4}
g(x^2) = \frac{1}{(1+x^2)^{3/2}} \;.
\end{equation}
We get
\begin{equation}\label{Jg_example_4}
\int\limits_0^\infty \frac{x}{(1+x^2)^{3/2}} \,J_0(w x) \,dx =
e^{-w} .
\end{equation}
On the other hand, according to \eqref{Jg_contour_II}, $J_g(w)$ is
given by the following sum of non-zero contour integrals
\begin{equation}\label{example_4}
J_g(w) = -\frac{1}{2}\lim_{r \rightarrow 0}(I_{C_3} + I_{C_4} +
I_{C_5}) \;,
\end{equation}
where the contours $C_i$ ($i=3,4,5$) are shown in
Fig.~\ref{fig:contour}. Since $I_{C_3} = I_{C_5}$ (see comments
after Eq.~\eqref{example_3}), changing variables $z = u e^{i\pi/2}$,
we obtain
\begin{align}\label{IC3+IC5}
I_{C_3} + I_{C_5} &= \frac{4}{\pi} \int\limits_{1+r}^\infty
\!\frac{u}{(u^2-1)^{3/2}} \, K_0(w u) \,du  \nonumber \\
&= \frac{4}{\pi} \frac{1}{\sqrt{r(2+r)}} \,K_0[w(1+r)] -
\frac{4w}{\pi} \int\limits_{1+r}^\infty \!\frac{u}{\sqrt{u^2-1}} \,
K_1(w u) \,du\;.
\end{align}
To calculate the contour integral $I_{C_4}$, we put $u = 1 - r
e^{i\phi}$, then
\begin{align}\label{IC4}
I_{C_4} &= \frac{2}{\pi\sqrt{r}} \int\limits_{\pi}^{-\pi} \!d\phi
\,e^{-i\phi/2} \frac{1 - r e^{i\phi}}{(2 + r e^{i\phi})^{3/2}} \,
K_0[w(1 - r e^{i\phi})] \big|_{r \rightarrow 0} \nonumber \\
&= - \frac{4}{\pi} \frac{1}{\sqrt{2\,r}} \, K_0(w) \left[ 1 +
\mathrm{O}(r) \right] \;.
\end{align}
As a result, we find in the limit $r \rightarrow 0$
\begin{equation}\label{example_4_final}
J_g(w) =  \frac{2 w}{\pi} \int\limits_1^\infty
\!\frac{u}{\sqrt{u^2-1}} \, K_1(w u) \,du = e^{-w} .
\end{equation}
Thus, Eq.~\eqref{Jg_example_4} is reproduced. It is in full
agreement with Corollary~2.

\begin{figure}[htb]
\begin{center}
\includegraphics[width=10cm]{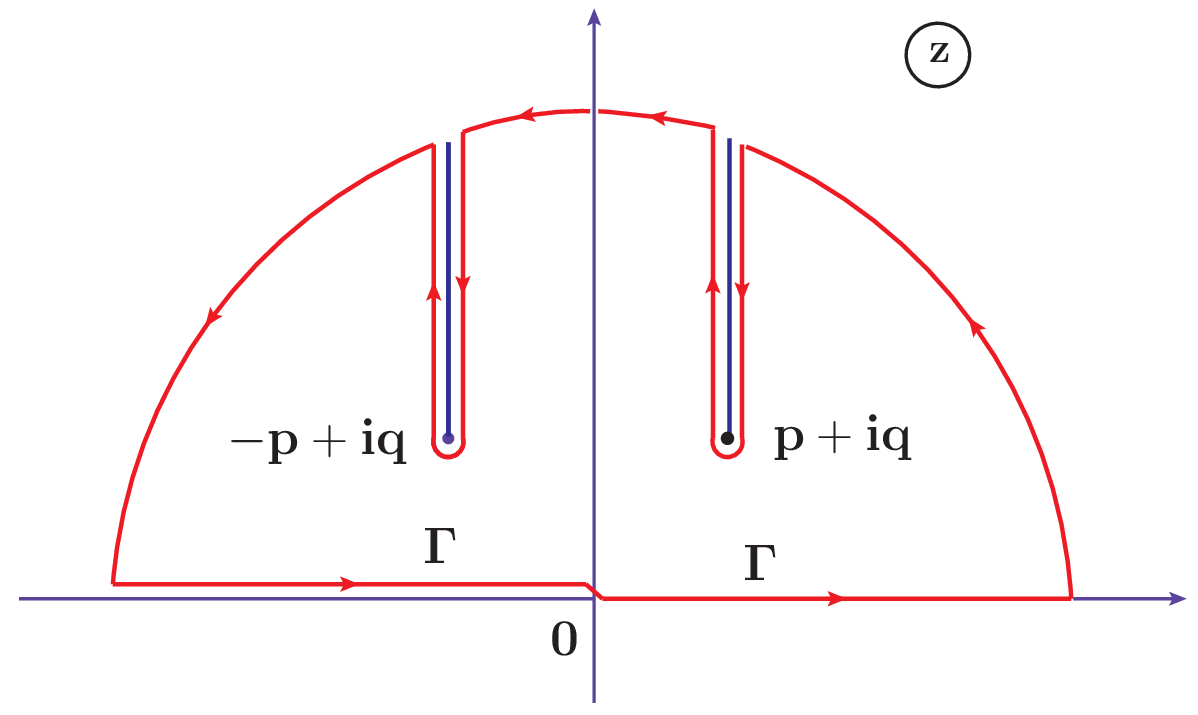}
\caption{The contour of integration for function \eqref{g_example_5}
with conditions $0 < a < b$.}
\label{fig:contour_5}
\end{center}
\end{figure}

\textsc{Example~5}. Consider the function
\begin{equation}\label{g_example_5}
g(x^2) = \frac{1}{\sqrt{x^4 + 2 a^2 x^2 + b^4}} \;,
\end{equation}
where $0 < a < b$. We have (see Eq.~2.12.6.11 in
\cite{Prudnikov_vol_2})
\begin{equation}\label{Jg_example_5}
\int\limits_0^\infty \!\frac{x}{\sqrt{x^4 + 2 a^2 x^2 + b^4}}
\,J_0(w x) \,dx = J_0 \!\left( \!w \sqrt{\frac{b^2 - a^2}{2}}
\right) \!K_0 \!\left( \!w \sqrt{\frac{a^2 + b^2}{2}} \right) .
\end{equation}
The function \eqref{g_example_5} has two branch cuts in the upper
plane, $(-p + iq, -p + i\infty)$ and  $(p + iq, p + i\infty)$, where
\begin{equation}\label{u_v_example_5}
p = \sqrt{\frac{b^2 - a^2}{2}} \;, \quad q = \sqrt{\frac{b^2 +
a^2}{2}} \;.
\end{equation}
The contour of integration is shown in Fig.~\ref{fig:contour_5}.
%

The contribution to $J_g(w)$ from the branch cut $(-p + iq, -p +
i\infty)$  is given by
\begin{equation}\label{Jg_5_1}
J_g^{(1)}(w) =  \frac{2i}{\pi} \int\limits_{q + ip}^{\infty +ip}
\!\frac{u}{\sqrt{[(u - q)^2 + p^2][(u + q)^2 + p^2]}} \,K_0(w u)
\,du  \;.
\end{equation}
Let us change variables $u = q + ip + z$ ($0 < z < \infty$). In
order to calculate an asymptotics of $J_g^{(1)}(w)$ as $w
\rightarrow \infty$, it is enough to put in \eqref{Jg_5_1} the
asymptotics
\begin{equation}\label{asym_K0_2}
K_0 \big( w (q+ip+z) \big) \big|_{w \gg 1} = \sqrt{\frac{\pi}{2w
(q+ip)}} \,e^{-w(q + ip - z)}  \left[ 1 + \mathrm{O}(w^{-1})
\right],
\end{equation}
and estimate a rest of the integrand in \eqref{Jg_5_1} near the
branch point $z=0$
\begin{equation}\label{integrand_5}
\frac{u}{\sqrt{[(u - q)^2 + p^2][(u + q)^2 + p^2]}} \bigg|_{z
\rightarrow 0} \simeq \frac{1}{2\sqrt{z}} \,\sqrt{\frac{q + ip}{2i p
q}} \;.
\end{equation}
Then we get
\begin{align}\label{Jg_1_exaple_5}
J_g^{(1)}(w)\big|_{w \gg 1} &=  \frac{1}{2\sqrt{\pi p q w}} \, e^{i
\pi/4} \,e^{-w(q + ip )} \int\limits_0^\infty \frac{dz}{\sqrt{z}}
\,e^{- z
w} \nonumber \\
&=  \frac{1}{2 w \sqrt{p q}} \, e^{- w q} \,e^{- i w(p - \pi/4)} \;.
\end{align}

Analogously, the contribution to the main asymptotics of $J_g(w)$
from the branch cut $(p + iq, p + i\infty)$ is equal to
\begin{equation}\label{Jg_2_exaple_5}
J_g^{(2)}(w)\big|_{w \gg 1} = \frac{1}{2 w \sqrt{p q}} \, e^{- w q}
\,e^{i w(p - \pi/4)} \;,
\end{equation}
that results in
\begin{equation}\label{asym_6}
J_g(w)\big|_{w \gg 1} = \frac{1}{w\sqrt{pq}} \cos \!\left[ w \left(p
- \frac{\pi}{4}\right) \right] e^{-q w} \,[1 + \mathrm{O}(w^{-1})]
\;.
\end{equation}
Note that we can derive the same asymptotics from the exact formula
\eqref{Jg_example_5}, if we use Eq.~\eqref{asym_Knu} and the
asymptotics of the Bessel function \cite{Bateman_vol_2}
\begin{align}\label{asym_Jnu}
J_\nu(z) \big|_{z \gg 1} &= \sqrt{\frac{2}{\pi z}}\,\cos \!\left(
\frac{4z - 2\nu\pi - \pi}{4} \right) \!\left[ 1 + \mathrm{O}(z^{-1})
\right] , \nonumber \\
&-\pi < \arg z < \pi \;.
\end{align}

Assume that, under conditions of Theorem~2, $k \geqslant 2$, and
real parts of one pair of the branch points coincide, $a_n=a_m=a$,
while $b_n < b_m$. Then our formula \eqref{Jg_contour_II} takes the
form
\begin{equation}\label{Jg_contour_II_mod}
J_g(w) = -\frac{1}{2} \left[ \sum_{
\begin{subarray}{c}
i=1 \\
i \neq n, m
\end{subarray}
}^k \,\int\limits_{C_i} \!\!z g(z^2) H_0^{(1)}(wz) \,dz +
\int\limits_{C_{nm}}\!\!\!z g(z^2) H_0^{(1)}(wz) \,dz \right] ,
\end{equation}
where $C_{nm}$ is a closed clockwise contour around the
finite-length branch cut $(a + ib_n, a + ib_m)$. This formula is
easily generalized to a case when real parts of two or more pairs of
branch points coincide. It is worth to consider a corresponding
example.

\begin{figure}[htb]
\begin{center}
\includegraphics[width=10cm]{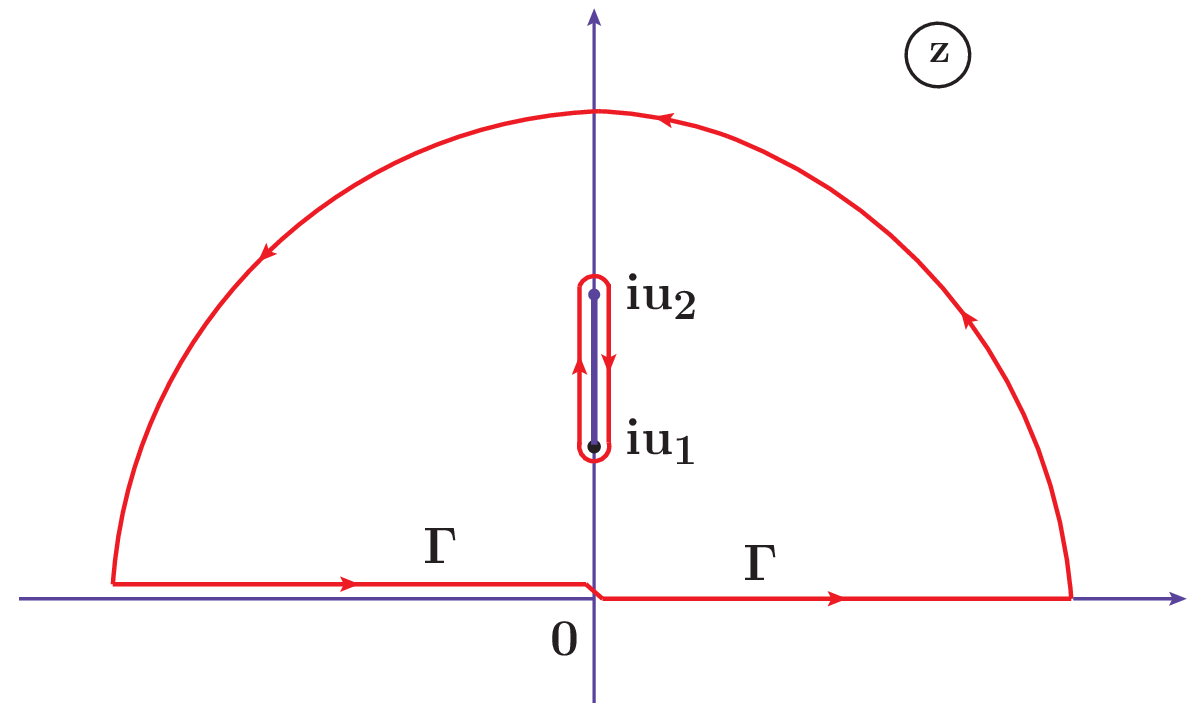}
\caption{The contour of integration for function \eqref{g_example_5}
with conditions $0 < b < a$.}
\label{fig:contour_6}
\end{center}
\end{figure}

\textsc{Example~6}. Let us take again function \eqref{g_example_5},
but with the other conditions $0 < b < a$.  In such a case, $g(z^2)$
is an analytic function in the upper half-plane with the branch cut
$(iu_1, iu_2)$, where
\begin{equation}\label{branch_cut_6}
u_1 = \sqrt{\frac{a^2 + b^2}{2}} - \sqrt{\frac{a^2 - b^2}{2}} \;,
\quad u_2 = \sqrt{\frac{a^2 + b^2}{2}} + \sqrt{\frac{a^2 - b^2}{2}}
\;,
\end{equation}
see Fig.~\ref{fig:contour_6}.
%
Starting from \eqref{Jg_contour_II_mod}, we come to the following
integral (see Eq.~2.16.4.1 in \cite{Prudnikov_vol_2})
\begin{align}\label{int_6}
J_g(w) &= \frac{2}{\pi} \int\limits_{u_1}^{u_2}
\!\frac{u}{\sqrt{(u_2^2 - u^2)(u^2 - u_1^2)}} \,K_0(w u) \,du \;,
\nonumber \\
&= I_0 \!\left( \frac{w(u_2 - u_1)}{2} \right) \!K_0 \!\left(
\frac{w(u_2 + u_1)}{2} \right) ,
\end{align}
in agreement with Eq.~2.12.6.10 in \cite{Prudnikov_vol_2}.

\section{Numerical evaluation of total cross section} %
\label{sec:num}

In this section we numerically calculate the total cross section for
a variety of expressions taken for the Born amplitude
$A_{\mathrm{B}}(s,t)$. Only Born amplitudes are considered that lead
to the correct asymptotics of the eikonal function at large $b$
\eqref{eik_bound}. The main goal is to verify whether our formula
approximates $\sigma_{\mathrm{tot}}$ well or not.

In what follows, in order to deal with dimensionless quantities in
numerical calculations, it is assumed that the momentum transfer $q
= \sqrt{-t}$ and collision energy $\sqrt{s}$ are measured in units
of $m_0$, where $m_0$ has a dimension of mass. Correspondingly, the
impact parameter $b$ is measured in units of $m_0^{-1}$. We
calculate the total cross section at some fixed energy, knowing the
Born amplitude at the same $s$.

\begin{figure}[htb]
\begin{center}
\includegraphics[width=8cm]{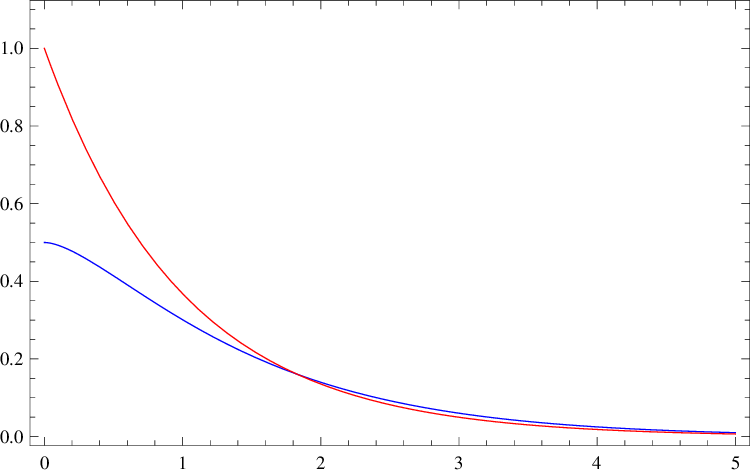}
\caption{Functions $e^{-x}$ (upper curve) and $x K_1(x)/2$ (lower
curve).} \label{fig:K1Exp}
\end{center}
\end{figure}

\textbf{1.} We start from the case when the Born amplitude is a
\emph{pure imaginary}, and it looks like
\begin{equation}\label{ImABorn_1}
\mathrm{Re}A_{\mathrm{B}}  = 0\;, \quad \mathrm{Im} A_{\mathrm{B}} =
\frac{4\pi a s}{(1 - t)^2} \;,
\end{equation}
where $a$ is a $t$-independent quantity to be defined below. It is
implied that $a$ may depend on $s$. The imaginary part of the Born
amplitude \eqref{ImABorn_1} is a meromorphic function of $q$ with
the double pole at $q=i$ in the upper half-plane. It corresponds to
the following eikonal function
\begin{equation}\label{ImEik_1}
\mathrm{Re}\chi(b) = 0\;, \quad \mathrm{Im} \chi(b) = a
\!\int\limits_0^{\infty} \! dq q \, J_0(q b) \frac{1}{(1 + q^2)^2} =
\frac{a}{2} \,b K_1(b) \;.
\end{equation}
The modified Bessel function $K_1(x)$ satisfies condition $\lim_{z
\rightarrow 0} [zK_1(z)] = 1$, and it has asymptotics
\eqref{asym_Knu}. The function $x K_1(x)/2$ is shown in
Fig.~\ref{fig:K1Exp} along with the function $e^{-x}$. Note that
\eqref{ImEik_1} is in full agreement with Corollary~1 of Theorem~1
(see also Example~2).

As it was mentioned above, $b$ is measured in units of $m_0^{-1}$.
It means that in fact
\begin{equation}\label{ImEik_1_dim}
\mathrm{Im} \chi(b) =  \frac{a}{2} \,m_0 \mathbf{b} K_1(m_0
\mathbf{b}) \;,
\end{equation}
where $\textbf{b}$ is the true impact parameter with a dimension of
(mass)$^{-1}$. If  $m_0$ depends on $s$, the eikonal function has no
factorized form. Its asymptotics is $\sim \exp(-m_0
\mathbf{b})/\sqrt{m_0 \mathbf{b}}$.
%

In the case under consideration, the total cross section is given by
\begin{equation}\label{cs_1}
\sigma_{\mathrm{tot}} = 4\pi \!\int\limits_0^{\infty} \! db b
\left\{ 1 - \exp \left[ -(a b/2) K_1(b) \right] \right\} .
\end{equation}
We also have
\begin{equation}\label{sigma1_1-2_def}
\sigma_1 = 4\pi a  \;, \quad \sigma_2 = -\frac{1}{3}\,\pi a^2  \;,
\end{equation}
\begin{equation}\label{sigma2_3_def}
\sigma_3 = \frac{2\pi}{3} \,a^3 \!\int\limits_0^{\infty} \!\!dx
\frac{x}{(1+x^2)^{2}} \!\int\limits_0^{\infty} \!\! dy
\frac{y}{(1+y^2)^{2}} \!\int\limits_0^{\infty} \!\! dz
\frac{z}{(1+z^2)^{2}} \, F_3(x,y,z) \;,
\end{equation}
where $F_3(x,y,z)$ is defined by eq.~\eqref{three_Bessels}.

Let $a=1/4$, then $\mathrm{Im}\chi(b) \leqslant 1/8$ for all $b$
(see Fig.~\ref{fig:K1Exp}). We obtain
\begin{equation}\label{cs_1.1}
\sigma_{\mathrm{tot}} = 0.979627 \,\pi \;,
\end{equation}
\begin{equation}\label{cs_1.2_1-2}
\sigma_1 + \sigma_2 = \frac{47}{48} \, \pi \;.
\end{equation}
Eq.~\eqref{sigma_3} gives us
\begin{equation}\label{cs_1.2_3}
\sigma_3 =  0.0014746 \;,
\end{equation}
and, correspondingly, $\sigma_1 + \sigma_2 + \sigma_3 = 3.077615$.

Let us define a relative error
\begin{equation}\label{error_def}
\varepsilon^{(n)} = \frac{|\sigma_{\mathrm{tot}} - \sum_{i=1}^n
\sigma_i|}{\sigma_{\mathrm{tot}}} \;.
\end{equation}
Then we find
\begin{equation}\label{eps_1.2}
\varepsilon^{(2)} = 4.7 \times 10^{-4} \;, \quad \varepsilon^{(3)} =
8.3 \times 10^{-6} \;.
\end{equation}
As one can see from \eqref{eps_1.2}, for moderately small eikonal
function ($|\chi(b)| \lesssim 0.1$) the accuracy of our approximate
formula \eqref{cs_expansion} is high even if one takes into account
only \emph{two first terms} in it.

\textbf{2.} Let us take a \emph{pure real} Born amplitude
\begin{equation}\label{ImABorn_2}
\mathrm{Re}A_{\mathrm{B}} = \frac{\pi s}{(1 - t)^2} \;, \quad
\mathrm{Im} A_{\mathrm{B}} = 0 \;.
\end{equation}
It corresponds to the eikonal function
\begin{equation}\label{ImEik_2}
\mathrm{Re}\chi(b) =  b K_1(b)/8 \;, \quad \mathrm{Im}\chi(b) = 0\;.
\end{equation}
Then calculations give us
\begin{equation}\label{cs_2}
\sigma_{\mathrm{tot}} = 0.0208237 \,\pi \;.
\end{equation}
The terms $\sigma_n$ with \emph{odd} $n$ are equal to zero,
$\sigma_1 = \sigma_3 = \sigma_5 = \ldots = 0$, while
\begin{equation}\label{sigma2_2}
\sigma_2 = \frac{\pi}{48} \;,
\end{equation}
\begin{align}\label{sigma2_4_def}
\sigma_4 &=  -\frac{\pi}{3 \cdot 2^9} \!\int\limits_0^{\infty}
\!\!dx \frac{x}{(1+x^2)^{2}} \!\int\limits_0^{\infty} \!\! dy
\frac{y}{(1+y^2)^{2}} \!\int\limits_0^{\infty} \!\! dz
\frac{z}{(1+z^2)^{2}} \!\int\limits_0^{\infty} \!\! du
\frac{u}{(1+u^2)^{2}}
\nonumber \\
&\times F_4(x,y,z,u) \;.
\end{align}
To make numerical estimates, we use formula \eqref{four_Bessels} and
find that
\begin{equation}\label{eps_2}
\varepsilon^{(2)} = 1.7 \times 10^{-4} \;, \quad \varepsilon^{(4)} =
5.0 \times 10^{-5} \;.
\end{equation}
It is a surprise that the first nonzero term in expansion
\eqref{cs_expansion} results in such a small relative error of
$\mathrm{O}(10^{-4})$.

\textbf{3.} Let us study a case when \emph{both real and imaginary}
parts of the Born amplitude are non-zero
\begin{equation}\label{ImABorn_3}
\mathrm{Re}A_{\mathrm{B}} = \mathrm{Im} A_{\mathrm{B}} =  \frac{\pi
s}{(1 - t)^2} \;.
\end{equation}
Correspondingly,
\begin{equation}\label{ImEik_3}
\mathrm{Re}\chi(b) = \mathrm{Im}\chi(b) =  b K_1(b)/8 \;.
\end{equation}
We get the numbers
\begin{equation}\label{cs_3}
\sigma_{\mathrm{tot}} = 0.999098 \,\pi \;,
\end{equation}
\begin{equation}\label{sigma3_1}
\sigma_1 = \pi = 3.141593 \;, \quad \sigma_3 = -0.00294914 \;.
\end{equation}
Note that $\sigma_2 = 0$ (see Eq.~\eqref{sigma_2}). As a result, we
find
\begin{equation}\label{eps_3}
\varepsilon^{(1)} = 9.0 \times 10^{-4} \;, \quad \varepsilon^{(3)} =
3.7 \times 10^{-5} \;.
\end{equation}

\textbf{4.} Finally, let us study the case when the real part of the
Born amplitude is zero, while its imaginary part is an entire
function of variable $q$
\begin{equation}\label{ImABorn_6}
\mathrm{Re}A_{\mathrm{B}} = 0 \;, \quad \mathrm{Im} A_{\mathrm{B}} =
\pi s \,e^{t/2} \;.
\end{equation}
It corresponds to the eikonal function of the form
\begin{equation}\label{ImEik_6}
\mathrm{Re}\chi(b) = 0 \;, \quad \mathrm{Im}\chi(b) = \frac{1}{4}
\,e^{-b^2/2} \;.
\end{equation}
Let us remember that the impact parameter $b$ is measured in units
of $m_0^{-1}$ (see our comments in the beginning of this section).
It means that in fact
\begin{equation}\label{ImEik_6_dim}
\mathrm{Im}\chi(b) = \frac{1}{4} \,e^{- ( m_0 \mathbf{b})^2/2} \;,
\end{equation}
where $\textbf{b}$ is the true impact parameter. Since the parameter
$m_0$ may depend on $s$, this eikonal function, in general, isn't
factorized.

The asymptotics of $\mathrm{Im}\chi(b)$ is in accordance with
Eq.~\eqref{f_asymp_1}. Indeed, our case \eqref{ImABorn_6}
corresponds to $f(x^2) = e^{x^2/2}$, $w = b$ in \eqref{Frenzen_int}
and, consequently, to $\xi = 1/2$, $\eta=0$ in \eqref{f_condition},
$\alpha = 1/2$ in \eqref{alpha}.

We obtain that
\begin{equation}\label{cs_6}
\sigma_{\mathrm{tot}} = 0.940816 \,\pi \;,
\end{equation}
\begin{equation}\label{sigma6_1}
\sigma_1 = \pi \;, \quad \sigma_2 = - \frac{\pi}{48} \;, \quad
\sigma_3 = 0.010897 \;.
\end{equation}
Aa in the previous examples, we come to the small relative errors
\begin{equation}\label{eps_6}
\varepsilon^{(2)} = 3.5 \times 10^{-3} \;, \quad \varepsilon^{(3)} =
1.6 \times 10^{-4} \;.
\end{equation}

From the above-studied examples we may conclude that for the
moderately small eikonal function ($|\chi(b)| \lesssim 0.1$, for all
values of the impact parameter $b$), the account of the first three
non-zero terms in expansion \eqref{cs_expansion} results in the
relative error of $\mathrm{O}(10^{-5})$. Throughout the paper, the
notation $\varepsilon = \mathrm{O}(10^{-n})$ means that
$\varepsilon$ is a
term of the order $(-n)$, i.e. $10^{-n} \leqslant |\varepsilon| < 10^{1-n}$.%
\footnote{Don't confuse it with the Landau symbol used to describe
the asymptotic behavior of functions.}


\section{Conclusions} %

In the present paper the eikonal approximation for the total cross
section for the scattering of two unpolarized particles is studied.
We have derived the approximate formula in the case when the eikonal
function $\chi(s,b)$ is moderately small (see
Eqs.~\eqref{cs_expansion}--\eqref{sigma_2},
\eqref{sigma_3}--\eqref{sigma_6}). Namely, we have shown that the
total cross section is given by the series of multiple improper
integrals of the Born amplitude $A_B(s,q)$, $q=\sqrt{-t}$. The
advantage of our formula compared to the standard eikonal formulas
is that our integrals contain no rapidly oscillating Bessel
functions. This circumstance is very important, since integrals with
the products of the Bessel functions are very difficult to evaluate
numerically because of their poor convergence and oscillatory
nature.

At large fixed $s$, the eikonal function decreases exponentially as
$b\rightarrow \infty$ \eqref{eik_bound}. On the other hand, it is
given by zero order Hankel transform of the function $A_B(s,q)$
\eqref{eikonal}. We have proved two theorems which allow one to
relate large $b$ behavior of $\chi(s,b)$ with the analytical
properties of the Born amplitude (for details, see
Section~\ref{sec:Hankel}). Six examples of these theorems are
presented.

To check the efficiency of our formula, we applied it for the
numerical calculations of the total cross section for a number of
particular expressions of $A_B(s,q)$. Only those Born amplitudes
were selected which result in moderately small eikonal functions and
lead to the correct asymptotics of  $\chi(s,b)$ at large $b$
\eqref{eik_bound}, in accordance with the theorems presented in
Sections~\ref{sec:exp} and \ref{sec:Hankel}. Let us underline, we
didn't demand that the resulting eikonal functions should have a
factorized form.

The results obtained allow us to make the following conclusions.
Provided that $|\chi(s,b)| \lesssim 0.1$ for all $b$, the sum of the
first three non-zero terms in expansion \eqref{cs_expansion}
approximates the total cross section with the relative error of
$\mathrm{O} (10^{-5})$. Higher accuracy can be achieved, if more
terms in \eqref{cs_expansion} ($\sigma_4, \sigma_5, \ldots$) are
taken into account.

Thus, if the Born amplitude is known at some fixed energy $\sqrt{s}$
as a function of the momentum transfer $q$, then our formulas
\eqref{cs_expansion}--\eqref{sigma_2}, \eqref{sigma_3} enable one to
numerically calculate the total cross section for the same
$\sqrt{s}$ with the high accuracy.



\setcounter{equation}{0}
\renewcommand{\theequation}{A.\arabic{equation}}

\section*{Appendix A}
\label{app:A}

Here we present the other terms in expansion \eqref{cs_expansion}
\begin{align}\label{sigma_3}
\sigma_3(s) &= \frac{1}{2^5 \pi^2 s^3} \prod_{i=1}^3
\int\limits_0^{\infty} dq_i q_i \, F_3(q_1,q_2,q_3) \Bigg[
\frac{1}{3} \prod_{i=1}^3 \mathrm{Im}A_B(s,q_i)
\nonumber \\
&\ - \mathrm{Im}A_B(s,q_1) \prod_{i=2}^3 \mathrm{Re}A_B(s,q_i)
\Bigg] ,
\end{align}
\begin{align}\label{sigma_4}
\sigma_4(s) &= \frac{1}{2^8\pi^3 s^4} \prod_{i=1}^4
\int\limits_0^{\infty} dq_i q_i \, F_4(q_1,q_2,q_3,q_4) \Bigg[
\prod_{i=1}^2 \mathrm{Im}A_B(s,q_i) \prod_{i=3}^4
\mathrm{Re}A_B(s,q_i) \,
\nonumber \\
&- \frac{1}{6} \prod_{i=1}^4 \mathrm{Re}A_B(s,q_i) - \frac{1}{6}
\prod_{i=1}^4 \mathrm{Im}A_B(s,q_i) \Bigg] ,
\end{align}
\begin{align}\label{sigma_5}
\sigma_5(s) &= \frac{1}{3 \cdot 2^{10}\pi^4 s^5} \prod_{i=1}^5
\int\limits_0^{\infty} dq_i q_i \, F_5(q_1,q_2,q_3,q_4,q_5) \Bigg[
\frac{1}{10} \prod_{i=1}^5 \mathrm{Im}A_B(s,q_i)
\nonumber \\
&+ \frac{1}{2} \, \mathrm{Im}A_B(s,q_1) \, \prod_{i=2}^5
\mathrm{Re}A_B(s,q_i)
\nonumber \\
&- \prod_{i=1}^3 \mathrm{Im}A_B(s,q_i) \prod_{i=4}^5
\mathrm{Re}A_B(s,q_i) \Bigg] ,
\end{align}
\begin{align}\label{sigma_6}
\sigma_6(s) &= \frac{1}{3 \cdot 2^{14}\pi^5 s^6} \prod_{i=1}^6
\int\limits_0^{\infty} dq_i q_i \, F_6(q_1,q_2,q_3,q_4,q_5,q_6)
\Bigg[\! -\frac{1}{15} \prod_{i=1}^6 \mathrm{Im}A_B(s,q_i)
\nonumber \\
&+ \frac{1}{15} \prod_{i=1}^6 \mathrm{Re}A_B(s,q_i) - \prod_{i=1}^2
\mathrm{Im}A_B(s,q_i) \prod_{i=3}^6 \mathrm{Re}A_B(s,q_i)
\nonumber \\
&+ \prod_{i=1}^4 \mathrm{Im}A_B(s,q_i) \prod_{i=5}^6
\mathrm{Re}A_B(s,q_i) \Bigg] .
\end{align}
The explicit expressions for the functions $F_i$ ($i=3,4,5,6$) are
given by Eqs.~\eqref{three_Bessels}, \eqref{four_Bessels},
\eqref{five_Bessels} and \eqref{six_Bessels} in Appendix~B.



\setcounter{equation}{0}
\renewcommand{\theequation}{B.\arabic{equation}}

\section*{Appendix B}
\label{app:B}

Let us define the integral with the product of several Bessel
functions
\begin{equation}\label{n_Bessels_def}
F_n(a_1, \ldots, a_n) = \int\limits_0^{\infty} \!\! dx x
\prod_{k=1}^n J_0(a_kx) \;,
\end{equation}
where $a_k > 0$, $n \geqslant 2$, $k=1, \ldots, n$. The case $n=2$
is well-known (see Eq.~6.512.8. in \cite{Gradshteyn})%
\footnote{In Ref.~\cite{Kisselev:16} this formula was generalized
for the product of two Bessel functions $J_\nu$ with $\mathrm{Re}\,
\nu
> -1$, $|\arg ab| < \pi/2$.}
\begin{equation}\label{two_Bessels}
F_2(a,b) = \int\limits_0^{\infty} \!\! dx x J_0(ax) J_0(bx) =
\frac{1}{a} \, \delta (a - b) \;.
\end{equation}

The integral with the product of three Bessel functions $J_0(x)$
($a,b,c > 0$),
\begin{align}\label{three_Bessels_def}
F_3(a,b,c) = \int\limits_0^{\infty} \!\! dx x J_0(ax) J_0(bx)
J_0(cx) \;,
\end{align}
is given by the formula (see Eq.~13.46(3) in \cite{Watson}, as well
as Eqs.~2.12.42.14., 2.12.42.15. in \cite{Prudnikov_vol_2})
\begin{equation}\label{three_Bessels}
F_3(a,b,c) = \left\{
  \begin{array}{cc}
   \displaystyle{\frac{1}{2\pi \Delta_3}} , & \Delta_3^2 > 0 \;, \\ \\
   0, & \Delta_3^2 < 0 \;,
  \end{array}
\right.
\end{equation}
where
\begin{equation}\label{Delta_3}
16 \Delta_3^2 = [c^2 - (a-b)^2][(a+b)^2 -c^2] \;.
\end{equation}
Note that expression \eqref{three_Bessels} is divergent if
$\Delta_3^2 = 0$ (it takes place when one of the parameters is equal
to the sum of the others, say, $c=a+b > 0$).

The complete expression for $F_4(a,b,c,d)$ was derived in
\cite{Kisselev:16}.%
\footnote{The corresponding formula in \cite{Prudnikov_vol_2} does
not include an important case $\Delta_4^2 < 0$.}
\begin{equation}\label{four_Bessels}
F_4(a,b,c,d) = \left\{
  \begin{array}{cl}
   \displaystyle{\frac{1}{\pi^2 \Delta_4} \, \mathrm{K} \!\left( \frac{\sqrt{abcd}}{\Delta_4} \right)},
   & \Delta_4^2 > abcd \;,
\\ \\
   \displaystyle{\frac{1}{\pi^2 \sqrt{abcd}} \, \mathrm{K} \!\left( \frac{\Delta_4}{\sqrt{abcd}} \right)},
   & 0 < \Delta_4^2 < abcd
\;,
\\ \\
   \displaystyle{\frac{1}{2\pi \sqrt{abcd}}} \, & \Delta_4^2 = 0 \;,
\\ \\
   0, & \Delta_4^2 < 0 \;,
  \end{array}
\right.
\end{equation}
where $a,b,c,d > 0$,
\begin{equation}\label{Delta_4}
16 \Delta_4^2 = (a + b + c - d)(a + b + d - c)(a + c + d - b)(b + c
+ d - a) \;,
\end{equation}
and
\begin{equation}\label{elliptic_integral}
\mathrm{K}(k) =  \int\limits_0^1 \! \frac{dt}{\sqrt{(1 - t^2)(1 -
k^2 t^2)}}
\end{equation}
is the complete elliptic integral of the first kind ($0 \leqslant k
< 1$) \cite{Bateman_vol_3}. The quantity $F_4(a,b,c,d)$
\eqref{four_Bessels} is not defined if $\Delta_4^2 = abcd$, since
$\mathrm{K}(k)$ diverges as $k \rightarrow 1$ \cite{Bateman_vol_3}
\begin{equation}\label{K_asymptotic}
\mathrm{K}(k)\big|_{k \rightarrow 1} \simeq \ln (4/\sqrt{1-k^2})\;.
\end{equation}
Nevertheless, since $F_4(a,b,c,d)$ will be used only as a part of
integrands, we can safely put $F_4(a,b,c,d) = 0$ for $\Delta_4^2 =
abcd$ in numerical calculations.

Finally, the following expressions were obtained in
\cite{Kisselev:16} ($a,b,c,d,e,f > 0$)
\begin{align}
F_5(a, b, c, d, e) &= \int\limits_0^{\infty} \!dt \,t \, F_3(a, b,
t) \, F_4(c, d, e, t) \label{five_Bessels} \;,
\\
F_6(a, b, c, d, e, f) &= \int\limits_0^{\infty} \!dt \,t \,F_4(a, b,
c, t) \,F_4(d, e, f, t) \label{six_Bessels} \;,
\end{align}
where $F_3(a_1,a_2,a_3)$ and $F_4(a_1, a_2, a_3, a_4)$ are defined
by formulas \eqref{three_Bessels} and \eqref{four_Bessels},
respectively. By doing in the same way, one can express integral
\eqref{n_Bessels_def} for $n > 6$ and $a_k > 0$, $k=1, \ldots, n$,
as a $(n-3)$-dimensional integral of algebraic functions. Let us
note that $F_n(a_1, \ldots, a_n) = 0$ for $n \geqslant 2$, if $a_n >
a_1 + a_2 + \ldots a_{n-1}$.



\setcounter{equation}{0}
\renewcommand{\theequation}{C.\arabic{equation}}

\section*{Appendix C}
\label{app:C}

Here we give a simple proof of formula \eqref{Legendre_limit}. The
differential equation for the Legendre function of the first kind
$u(z) = P_\nu^{-\mu}(z)$ looks like \cite{Bateman_vol_1}
\begin{equation}\label{Legendre_equation}
(1 - z^2)\frac{d^2}{dz^2} \,u(z) -2z \frac{d}{dz} \,u(z) + \left[
\nu(\nu + 1) - \frac{\mu^2}{1 - z^2}\right] \!u(z) = 0 \;.
\end{equation}
Let us change variables
\begin{equation}\label{z_to_x}
z = \cos \frac{x}{\nu} \;.
\end{equation}
In the limit $\nu \gg 1 \gg x$, we get
\begin{equation}\label{dz+to_dx}
z \frac{d}{dz} \simeq - \frac{\nu^2}{x} \,\frac{d}{dx} \;, \quad (1
- z^2) \frac{d^2}{dz^2} \simeq \nu^2 \!\left[ \frac{d^2}{dx^2} -
\frac{1}{x} \frac{d}{dx} \right] ,
\end{equation}
and differential equation \eqref{Legendre_equation} takes the form
\begin{equation}\label{Bessel_equation}
x^2 \frac{d^2}{dx^2} \,u(x) + x \frac{d}{dx} \,u(x) + (x^2 - \mu^2)
u(x) = 0 \;,
\end{equation}
which is Bessel's equation \cite{Bateman_vol_2}. It means that
\begin{equation}\label{Legendre_via_Bessels}
\lim_{\nu \rightarrow \infty} \,P_\nu^{-\mu} \!\left( \cos
\frac{x}{\nu} \right) = c_1 J_\mu(x) + c_2 Y_\mu(x) \;,
\end{equation}
where $c_1, c_2$ are constants. The singular solution to the Bessel
equation $Y_\mu(x)$ is ruled out, since $P_\nu^{-\mu} \!\left( \cos
\frac{x}{\nu} \right)$ is regular at $x=0$. So, $c_2 = 0$. We obtain
\begin{equation}\label{Legendre_small_x}
P_\nu^{-\mu} \!\left( \cos \frac{x}{\nu} \right)\Big|_{x \ll 1 \ll
\nu} = \frac{1}{\Gamma(1 + \mu)} \left( \frac{x}{2} \right)^{\!\mu}
= J_\mu(x) \Big|_{x \ll 1} \;,
\end{equation}
that results in $c_1 = 1$, and Eq.~\eqref{Legendre_limit} is
reproduced.



\setcounter{equation}{0}
\renewcommand{\theequation}{D.\arabic{equation}}

\section*{Appendix D}
\label{app:D}

Let us prove the formula
\begin{equation}\label{diff_n}
\left(\frac{1}{z} \frac{d}{dz} \right)^{\!\!n} = \sum_{k=0}^{n-1}
\frac{(-1)^k}{2^k} \frac{\Gamma(n+k)}{\Gamma(k+1)\Gamma(n-k)}
\frac{1}{z^{n+k}} \frac{d^{n-k}}{dz^{n-k}} \;,
\end{equation}
where $n = 1,2,3, \ldots$, using mathematical induction. It is
evidently that our statement \eqref{diff_n} is valid for $n=1$. Let
us assume that it holds for some integer $n=m > 1$,
\begin{equation}\label{diff_m}
P(m) = \left(\frac{1}{z} \frac{d}{dz} \right)^{\!\!m} =
\sum_{k=0}^{m-1} \frac{(-1)^k}{2^k}
\frac{\Gamma(m+k)}{\Gamma(k+1)\Gamma(m-k)} \frac{1}{z^{m+k}}
\frac{d^{m-k}}{dz^{m-k}} \;.
\end{equation}
Then we find using Eq.~\eqref{diff_m}
\begin{align}\label{diff_m+1}
\left(\frac{1}{z} \frac{d}{dz} \right)^{\!\!m+1} &= \sum_{k=0}^{m-1}
\frac{(-1)^k}{2^k} \frac{\Gamma(m+k)}{\Gamma(k+1)\Gamma(m-k)}
\nonumber \\
&\times \left[ \frac{1}{z^{m+1+k}} \frac{d^{m+1-k}}{dz^{m+1-k}} -
(m+k) \frac{1}{z^{m+2+k}} \frac{d^{m-k}}{dz^{m-k}} \right] \nonumber \\
&= \sum_{k=0}^{m} \frac{(-1)^k}{2^k}
\frac{\Gamma(m+1+k)}{\Gamma(k+1)\Gamma(m+1-k)} \frac{1}{z^{m+1+k}}
\frac{d^{m+1-k}}{dz^{m+1-k}} \nonumber \\
&= P(m+1) \;.
\end{align}
So, the inductive case holds. Consequently, formula \eqref{diff_n}
is valid for all integer $n\geqslant 1$. \textsc{Q.E.D.}




\end{document}